\documentclass[12pt]{arXiv_class}

\title{\textbf{Fast Polarization Switch for Polarization-Based Quantum Communication}} 
\author{Vinicius M. Lima, Felipe Calliari, Joaquim D. Garcia, Jo\~{a}o Pedro D. Garcia, \\Gustavo C. Amaral, Guilherme P. Tempor\~{a}o, and Jean Pierre von der Weid}

\begin{document}
\maketitle

\begin{abstract}
We present a complete optoelectronic unit for polarization visualization, switching and control. The system is based on an FPGA unit and comprises: an acquisition unit containing an analog polarimeter and digital-to-analog converters; an FPGA capable of implementing an optimal algorithm for three-stage arbitrary polarization tracking; and an electronic driver with analog-to-digital converters capable of interfacing with Lithium-Niobate-based Polarization Controllers. The results, determined via simulation of real-parameter devices, show that fast polarization switch is achievable.
\end{abstract}

\section{Introduction}

Polarization stabilization and control is a matter of particular interest in the area of quantum communication mainly due to the possibility of encoding qubits in polarization states \cite{GisinRMP02}. Recently, quantum communication protocols based on two-photon interference have been given a lot of attention, specially with the advent of the Measurement Device Independent Quantum Key Distribution Protocol \cite{LoPRL2012}. This protocol, in particular, makes use of the photon-bunching effect in a Bell-State Projection (BSP) which is highly dependent on the indistinguishability between quantum states and, thus, on the aligned polarization states at the BSP \cite{ThiagoPRA2013}. Therefore, polarization stabilization along the quantum channel becomes an important issue.

\begin{figure}[htbp]
\centering
\includegraphics[width=0.8\linewidth]{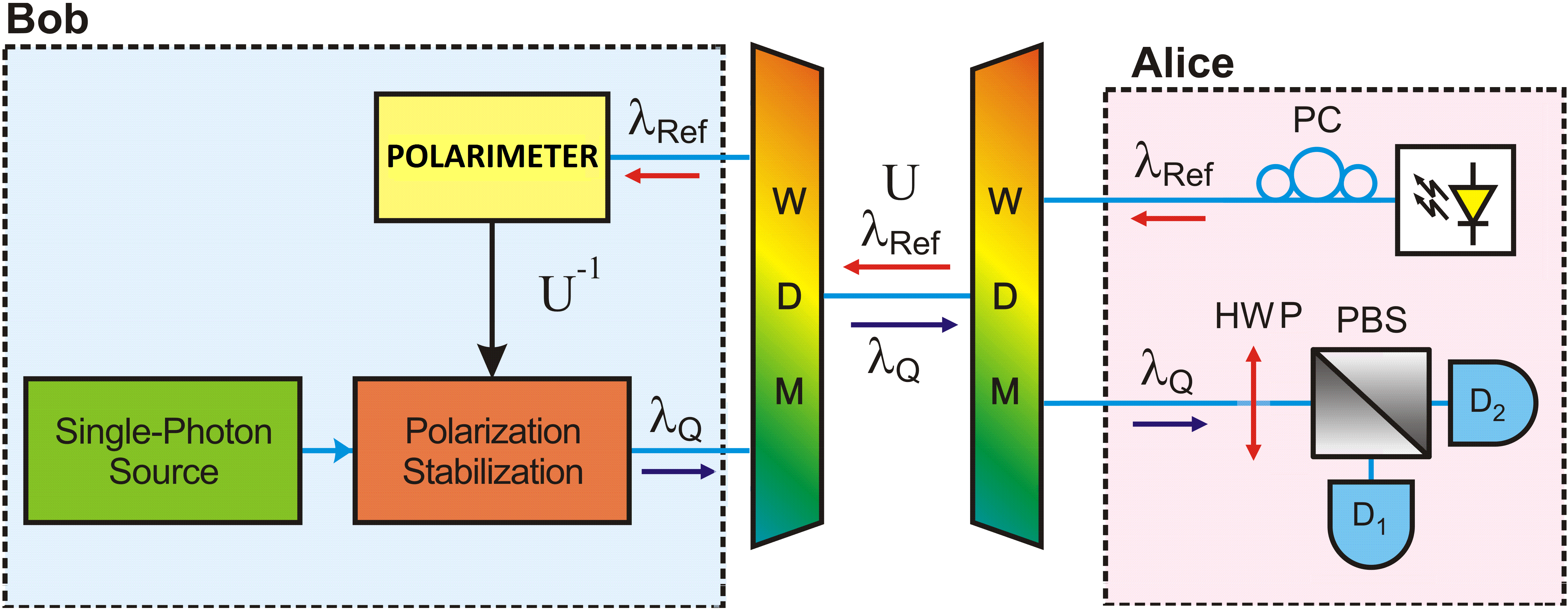}
\caption{Simplified setup for on-time polarization tracking in wavelength multiplexed optical quantum channels.}
\label{fig:PolarizationState_Stabilizer}
\end{figure}

Recently, a scheme for active polarization stabilization along the optical quantum channel employing current optical fibre technology has been successfully demonstrated \cite{XavierOPEX08,ThiagoQIM12}. Such scheme makes use of a reference laser which is launched in the fiber in a DWDM channel adjacent to that of the quantum communication channel in a counter-propagating manner, such that the transmitter receives the reference signal from the receiver \cite{XavierOPEX08}. A simplified scheme of the proposed system is depicted in Fig.\ref{fig:PolarizationState_Stabilizer}.

By rotating the received polarization state to match what should be its original state, the transmitter gains access to the optical link transfer function and can produce the inverse transformation on the outgoing optical pulses \cite{XavierOPEX08}. Not only the polarization has been shown to stabilize during long term communication periods but it was also proven that the noise contribution from the adjacent counter-propagating channels was negligible, enabling the use of the technique in quantum communication channels \cite{XavierOPEX08}.

However efficient, the polarization control enforced by the system described above only enables the stabilization of the polarization state rather than its fast switching. As the quantum states must be encoded in the polarization, the rapid switching between polarization states becomes a rather important task, even more when the mathematical security proofs for quantum communication channels rely on the rate of secret key generation \cite{MaPRA05,LutkenhausPRA00}. By associating the polarization stabilization setup proposed in \cite{XavierOPEX08} and an active feedback system based on a sample of the outgoing polarization state, a mixed analog-digital polarization switch is proposed. The scheme is supposed to rapidly switch between arbitrary polarization states whilst protecting the quantum state alignment from fluctuations of the optical channel, a claim supported by the numerical simulation results of each individual system that comprise the main unit.
	
The paper is divide as follows. In Section II, the mathematical formalism necessary to represent stats of polarization and the lithium-niobate-based polarization controller is presented. In Section III, the individual units are detailed and the complete optoelectronic polarization switch unit is presented. The simulation results and discussions appear in Section IV. Finally, in Section V, the conclusions are drawn and the real implementation of the envisioned system is discussed.

\section{Theoretical Formulation}

\subsection{Mathematical representation of polarization}\label{mathRep}

The state of polarization of light has been represented mathematically as Jones Vectors and Stokes Vectors \cite{saleh1991fundamentals}. Even though the conversion between these two representations require straightforward computations, we shall stick to the Stokes representation since visualization in the Poincar\'e Sphere is direct \cite{saleh1991fundamentals}. Stokes vectors are 4-dimensional vectors that carry information about the State of Polarization (SOP) of light. Since the first component ($S_0$) is associated to the total light intensity, it is common to normalize the Stokes Vector by dividing all of its entries by $S_0$. In the case of polarized light, the 3-dimensional Stokes Vector formed of the remaining three normalized components of the former 4-dimensional vector, has norm $1$. Indeed, the degree of polarization of light is given by the norm of the Stokes vector. Assuming that light is polarized enables one to simplify the representation. The usual representation of the 3-dimensional normalized Stokes vector, assuming polarized light (and, hence norm 1), is a point in the 3-sphere known as the Poincar\'{e} Sphere. Since all SOPs are mapped bijectively in the 3-sphere, we shall treat, from now on, an SOP as a point in the 3-sphere.

SOP changes that do not affect the light intensity can be represented by rotations in 3-space. These rotations are a class of unitary transformations and can be represented by orthonormal matrices \cite{strang09}. The rotation matrix in 3-space has a very interesting characterization via the Spectral Theorem: it always has 3 eigenvalues (since they are normal); all of the eigenvalues have norm 1; one of the eigenvalues is always equal to 1 and its eigenvector is the rotation axis, $e$; the remaining eigenvalues are complex conjugate numbers whose real part correspond to the cosine of the rotation angle, $\theta$, and whose imaginary part correspond to the sine of the rotation angle. A simple and robust way of manipulating 3-space rotation matrices is through the quaternions number system \cite{karlsson2004quaternion,DiasGarciaArXiV2016}. Despite the fact that polarisers affect the light intensity and, as such, cannot be represented by orthonormal matrices, they can be represented by projection matrices.

\subsection{Lithium-Niobate-based Polarization Controller Characteristics}\label{Control}

Literature around polarization control is very rich and diverse techniques were proposed and verified over the last years \cite{heismann1994analysis69, imai1985optical46, noe1988automatic40, heismann1989integrated55}. Our methodology focuses on the electro-optic $LiNbO_3$ EOSpace Polarization Controller Module (PCM) which is available commercially as a multi-stage component to which the input voltage may vary within a $\pm70$ Volts range \cite{saleh1991fundamentals}. 

A single stage of the PCM has 3 electrodes \cite{EOSpace} and realizes an arbitrary \textit{Linear Retarder}. A linear retarder is a linear wave-plate capable of inducing a relative phase difference between the two polarization axes. This accomplished through the bi-refringent characteristic of the wave-plate which cause two orthogonal polarization axes to experiment different indexes of refraction while traversing the material. The difference in propagation time is, thus, responsible for enforcing the relative phase shift between polarization axes. 

Linear Retarders have a main polarization axis, also known as eigen-mode, $e\in \{v\in\mathbb{R}^3| z=0\}$, and a characteristic phase delay, $\theta\in[0, 2\pi)$. It is possible to show that, by changing the eigen-mode and the phase delay of a linear retarder, one can shift from one SOP to any other SOP \cite{DiasGarciaArXiV2016}. In order to  set the eigen-mode to $e=(\cos(\alpha/2), \sin(\alpha/2),0)$ and the phase delay to $\theta = 2\pi\delta$ the electrodes voltages of the lithium-niobate-based PCM must be set to \cite{EOSpace}:
\begin{align}
&V_a=2 V_0 \delta \sin\pars{\alpha} - V_{\pi}\delta \cos\pars{\alpha}+V^b_{a}\label{ds_EOS1} \\
&V_b=0 \label{ds_EOS2} \\
&V_c=2 V_0 \delta \sin\pars{\alpha} - V_{\pi} \delta \cos\pars{\alpha}+V^b_{c}\label{ds_EOS3} 
\end{align}
where $V_{\pi}$ is the voltage required to induce a $180^o$ phase shift between the TE and TM modes for a single stage, $V_0$ is the voltage required to rotate all power from the TE to the TM mode, or vice versa, for a single stage, and $V^b_{a}$ and $V^b_{c}$ are the bias voltages required on electrodes A and C, respectively, in order to achieve zero bi-refringence between the TE and TM modes \cite{EOSpace}. Even though the data-sheet of the device provides the voltage range for $V_0$, $V_{\pi}$,$V^b_{a}$ and $V^b_{c}$, their actual values for an arbitrary stage must be determined via a calibration procedure.

Algorithms for both the calibration procedure and for identifying the required rotation in order to achieve a desired polarization state at the output of the polarization controller are presented in \cite{xi2010novel, DiasGarciaArXiV2016}. We shall focus on the hardware implementation of the control system.

\section{Optoelectronic Setup Proposal}

\begin{figure}[htbp]
\centering
\includegraphics[width=0.8\linewidth]{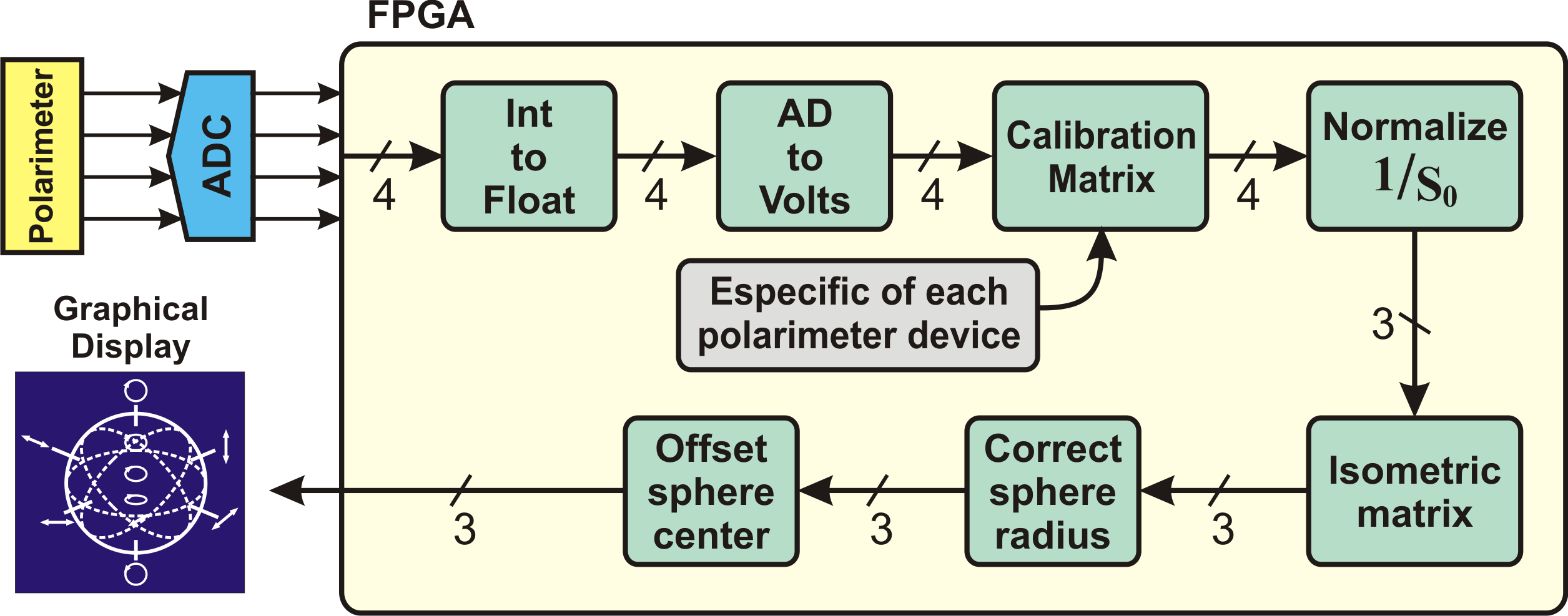}
\caption{Block diagram of the FPGA's internal structure for polarization visualization. The results acquired by the polarimeter are sent to the FPGA and the result is displayed graphically.}
\label{fig:PolarimeterSchematic}
\end{figure}

The representation of a polarization state in the Poincar\'{e} Sphere is achieved by measuring the Stokes vector of the incident light, which can be performed in a number of ways with either a set of optical splitters, wave plates and detectors or with a polarization rotator and a polarization state reference \cite{SalehTeichBOOK}. The proposed architecture makes use of the FPGA-based polarimeter for polarization state visualization proposed in \cite{CalliariTCC2014}. The FPGA's internal schematic is presented in Fig.\ref{fig:PolarimeterSchematic}.

The polarimeter measures the intensities on each of the three main polarization basis, the rectilinear, diagonal, and circular basis, from which one is capable of determining the corresponding Stokes vector. The measurement is represented by a voltage value which is sent to an Analog-to-Digital Converter (ADC) so the digitized values can be interpreted by the FPGA. Inside the FPGA structure, a series of sub-blocks perform the signal interpretation: the \textit{Int to Float} block is responsible for representing the digitized voltage values as floating point numbers; the \textit{AD to Volts} block converts the digitized values back to their original voltage values represented as a floating point number; the \textit{Calibration Matrix} block permits one to associate the voltage value measured by the polarimeter to each of the four un-normalized entries of the Stokes vector (this block is dependent on the device and on the operating wavelength); up to the \textit{Normalize} block, the digital structure deals with the 4-dimensional Stokes vector, but after they are normalized, we need only to deal with the 3-dimensional vector; the \textit{Isometric Matrix} block enables the 2-dimensional visualization of the Poincar\'{e} Sphere; the remaining \textit{Correct} and \textit{Offset} blocks associate the pixels in the graphic display with the values of the SOP vector.

\begin{figure}[htbp]
\centering
\includegraphics[width=0.8\linewidth]{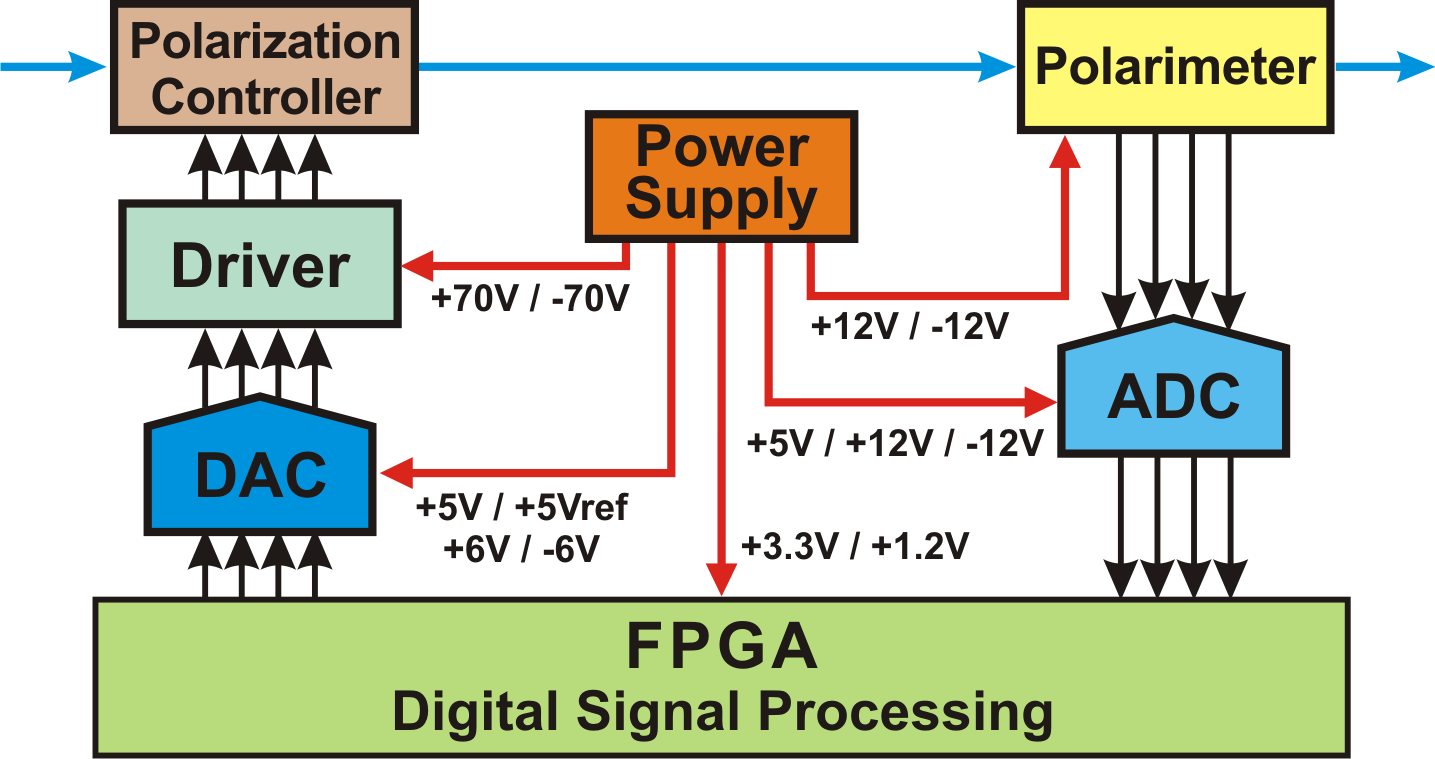}
\caption{Mixed analog-digital feedback loop for polarization visualization, selection, and stabilization.}
\label{fig:ElectronicSetup}
\end{figure}

The goal is to combine the setups of Figs. \ref{fig:PolarizationState_Stabilizer} and \ref{fig:PolarimeterSchematic} in a feedback loop enabling visualization and control of the output SOP \cite{DiasGarciaArXiV2016}. The electronic implementation of the feedback system, depicted in Fig. \ref{fig:ElectronicSetup}, is composed of: an analog polarimeter; an array of analog-to-digital converters (ADC); an FPGA unit; an array digital-to-analog converters (DAC); an electronic driver unit; a Lithium-Niobate-based PCM; and a power supply.

\begin{figure}[htbp]
\centering
\includegraphics[width=0.8\linewidth]{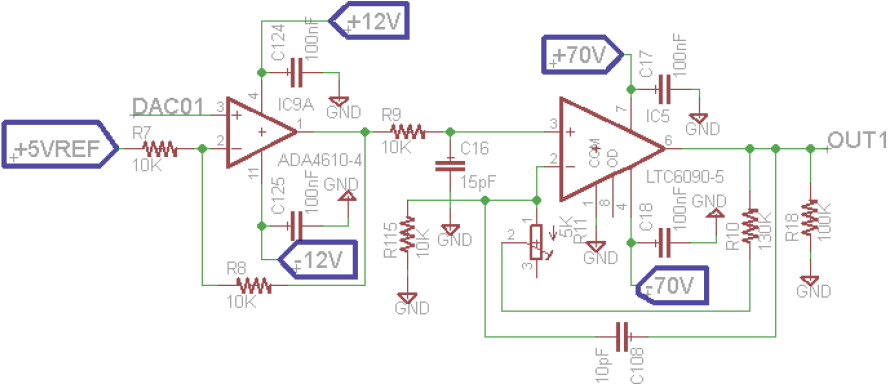}
\caption{Electronic schematic of a single $\pm70$ Volts driver stage of the lithium-niobate-based PCM.}
\label{fig:DriverSchematic}
\end{figure}

Running in parallel inside the FPGA is the algorithm developed in \cite{DiasGarciaArXiV2016}. The algorithm takes the current state of polarization and the target state of polarization as inputs and attempts to output voltage levels which perform a rotation (according to the equations describing the PCM behavioral) so that the current and target SOPs match each other. The electronic driver that follows the DAC guarantees that the low power signals sent from the FPGA are converted to match the Lithium-Niobate based PCM. Each stage of the PCM demands two electronic drivers capable of reaching an output voltage swing between $\pm70$ Volts for which the full electrical schematic of a single driver stage is presented in Fig. \ref{fig:DriverSchematic}.

\begin{figure}[htbp]
\centering
\includegraphics[width=0.8\linewidth]{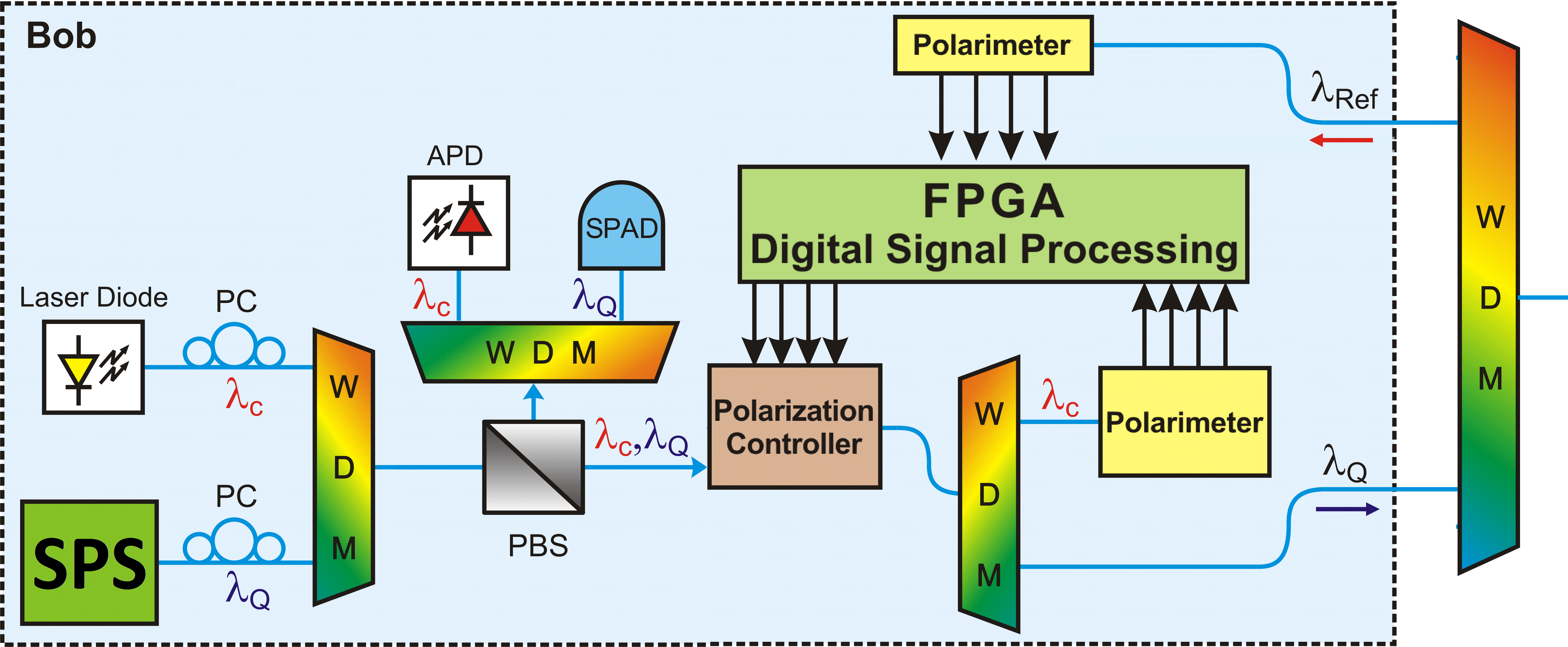}
\caption{Proposed full polarization stabilization and selection scheme for quantum communication channels. \textbf{SPS}: Single-Photon Source; \textbf{WDM}: Wavelength Division Multiplexer; \textbf{PBS}: Optical Polarizing Beam Splitter; \textbf{APD}: Avalanche Photodiode \textbf{SPAD}: Single-Photon Avalanche Photodiode; \textbf{PC}: Mechanical Polarization Controller.}
\label{fig:PolarizationState_SelStab_Quantum}
\end{figure}

The setup presented in Fig.\ref{fig:ElectronicSetup} is suited for classical communication channels since a sample of the output signal can be measured by the polarimeter. When one is interested in transmitting quantum states, as for quantum communication purposes, however, the state cannot be measured without being destroyed. Also, a sample of the quantum state cannot be taken. Thus, a different approach should be considered: the measured polarization state cannot be a sample of the outgoing quantum state so it must be a classical state. It should, however, be polarization-aligned with the outgoing quantum state so, by measuring the classical state, one gains information over the polarization state of the outgoing quantum state. If we suppose the quantum state is generated by an ideal single-photon source (SPS), the block diagram depicted in Fig. \ref{fig:PolarizationState_SelStab_Quantum} should be able to perform its polarization control.

In the input of the polarization control setup, a reference optical signal from a tunable laser diode is directed to a WDM combiner together with the output of the SPS. It is important that each source occupies different WDM channels so they can be multiplexed and later demultiplexed. A polarizing beam splitter (PBS) is then connected to the output of the WDM to act as a polarization aligner. By minimizing the outputs from detectors SPAD$_{adj}$ and APD$_{adj}$, the alignment of the polarization states of both channels at the remaining PBS output arm is enforced \cite{AmaralOL2016}. The polarization-aligned channels are then directed to a PCM where they experiment the same polarization rotation given the wavelengths do not differ greatly \cite{XavierOPEX08}. The polarization-rotated signals are divided by a WDM splitter so $\lambda_Q$ is directed to the output of Bob's station and $\lambda_C$ is directed to the mixed analog-digital feedback system. Combining the information of which polarization state should be keyed at the output and how the optical channel influences the polarization of light, the outgoing state is composed. Note that the quantum state is not measured at any time in the polarization control system and we infer its polarization state taking the measurement over the classical polarization-aligned state as a reference.

\section{Results}

\begin{figure}[htbp]
\centering
\includegraphics[width=0.8\linewidth]{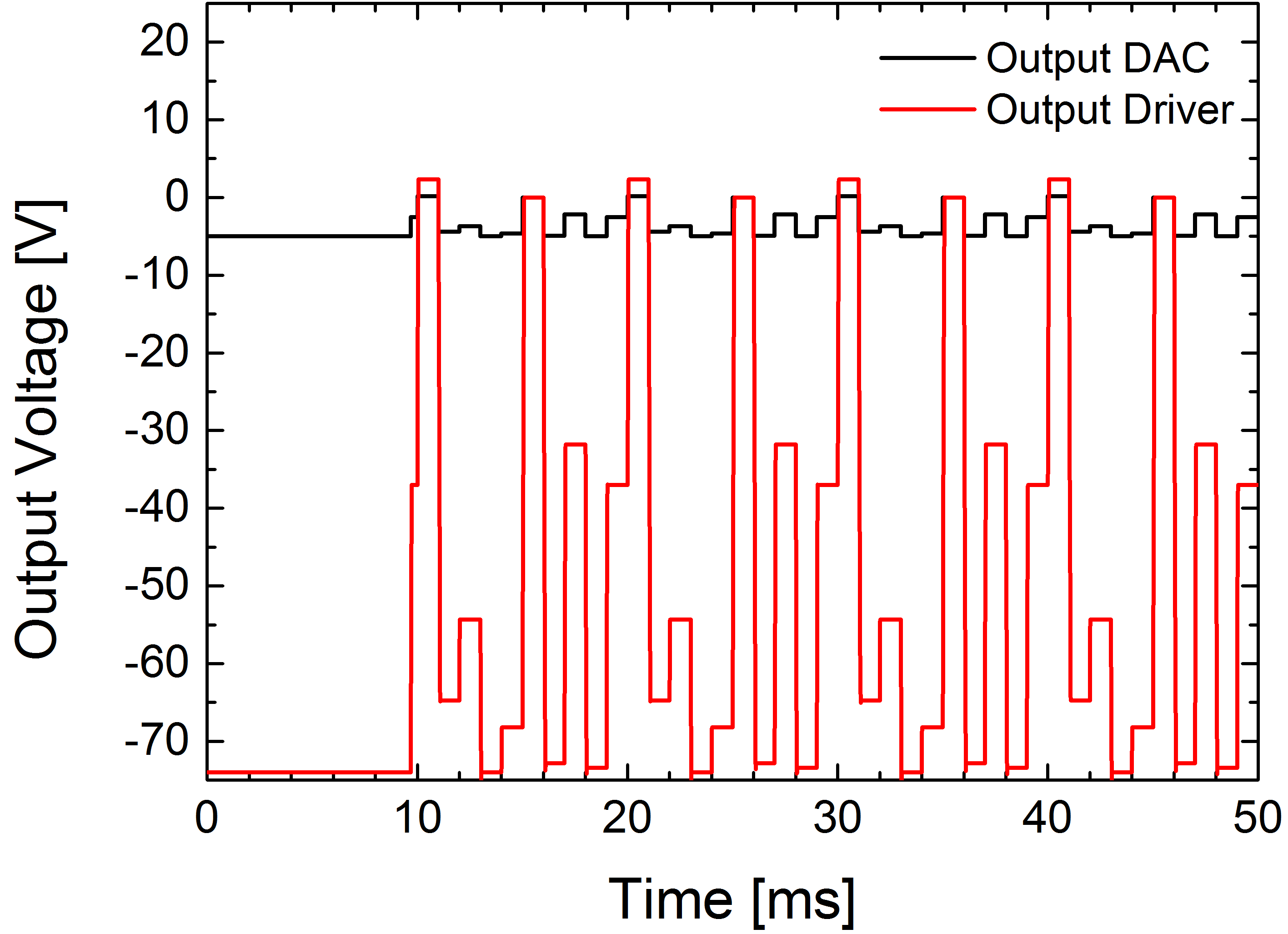}
\caption{Voltage swing at the driver output as a function of the DAC output. The FPGA is responsible for configuring the DAC output.}
\label{fig:OutputVariation}
\end{figure}

The simulation results were acquired using the Spice simulation tool. All the electronic devices were replaced by their real-parameter model which take into account the supply saturation, frequency response (slew-rate), and gain limitations. In Fig. \ref{fig:OutputVariation}, we depict the simulation results for various digital inputs at the DAC and the corresponding output from both the DAC and the power driver.

\begin{figure}[htbp]
\centering
\includegraphics[width=0.8\linewidth]{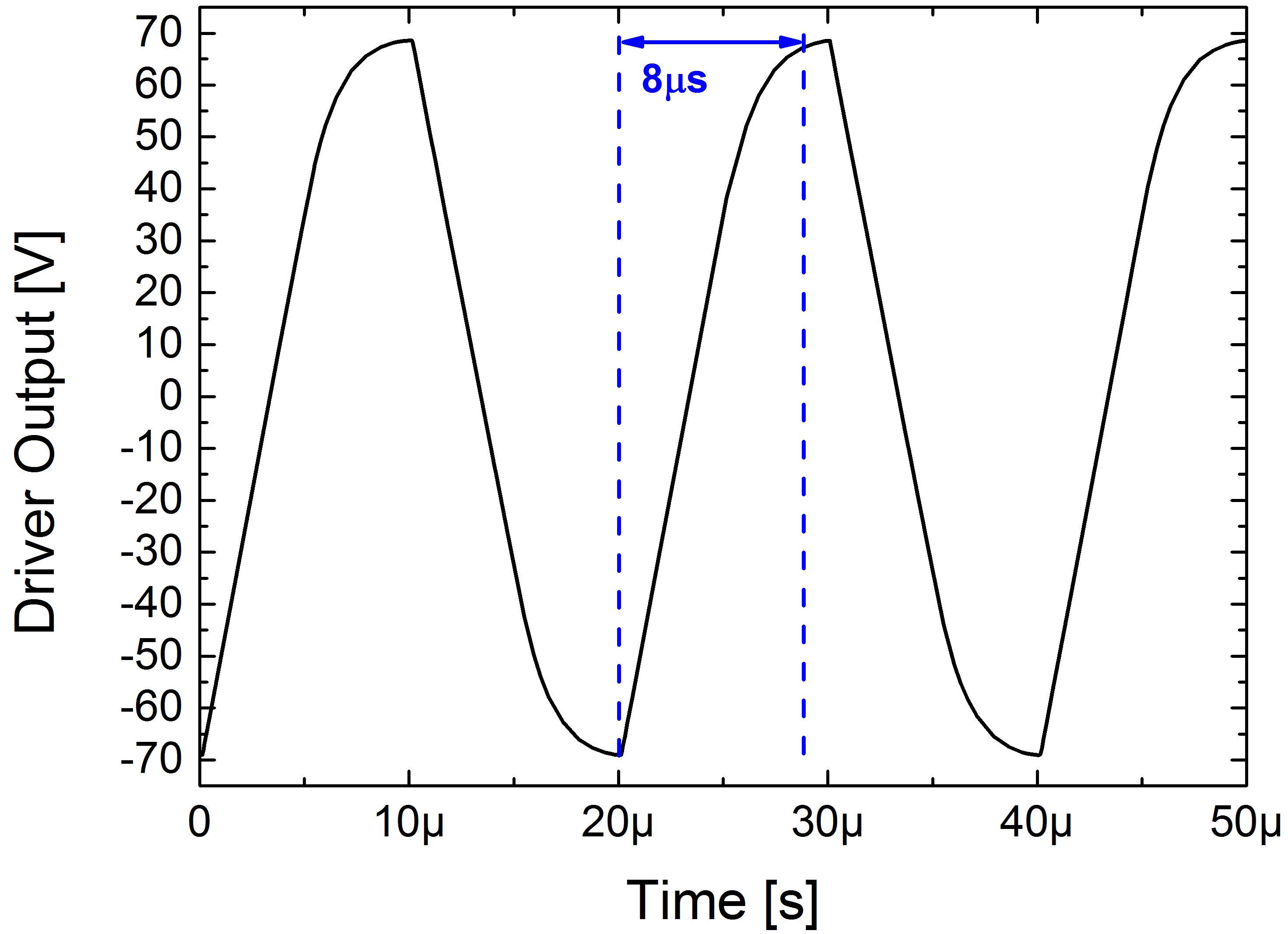}
\caption{Detailed transition time in the full $\pm70$ voltage swing at the electronic driver's output.}
\label{fig:TransitionTime}
\end{figure}

We observe that the voltage swing is respected by the power driver unit. An important information, however, is the one regarding the switching time between one voltage level to the other since, ultimately, it determines our maximum achievable transmission rate in polarization-encoded quantum communication channels. In Fig. \ref{fig:TransitionTime}, we detail the switching time and show that the proposed system is capable of achieving an up to $125$ KHz transmission rate. It should be noted, however, that the slew-rate of the LTC6090-5 is dependent on its gain. Throughout the simulation runs, we enforced a $14$V/V gain which limits the transition time to $8$kHz. By introducing a pre-amplifier stage with $3$V/V gain before the LTC6090-5, we could operate under a $5$V/V gain achieving a $1$MHz switching rate \cite{LTC6090}.

\section{Conclusion}\label{conclusion}

An optoelectronic unit comprised of a mixed analog-digital control loop and a polarization alignment state is presented. The system's application in polarization-based quantum communication links is studied with respect to the switching rate, usually above $1$MHz. Considering the device's frequency response, we show, through numerical simulations, that the demands are matched. We also show that polarization visualization can be performed throughout the switched operation of the system and a polarization visualization unit is proposed and experimentally demonstrated. Fast and embeddable signal processing tools that permit state of polarization control have been described in the literature and are directly implementable in the proposed scheme.

\section*{Acknowledgment}

The authors would like to thank brazilian agency CNPq for financial support.

\section*{Supplemental Material}

The authors provide digital supplemental material to accompany the article: a video file depicting the simulation run of the polarimeter can be accessed in \cite{YouTubeVideo_Polarimeter}.

\bibliographystyle{IEEEtran}
\bibliography{Ref}

\vfill

\end{document}